# A Dark Excited State of Fluorescent Protein Chromophores, Considered as Brooker Dyes

*Seth Olsen and Ross H. McKenzie*

Centre for Organic Photonics and Electronics, School of Mathematics and Physics, The University of Queensland, Brisbane, QLD 4072 Australia

s.olsen1@uq.edu.au

***Abstract*** The green fluorescent protein (GFP) chromophore is an asymmetric monomethine dye system. In the resonance color theory of dyes, a strong optical excitation arises from interactions of two valence-bond structures with a third, higher structure. We use correlated quantum chemistry to show that the anionic chromophore is a resonant Brooker dye, and that the third structure corresponds to a higher stationary electronic state of this species. The excitation energy of this state should be just below the first excitation energy of the neutral form. This has implications for excited state mechanism in GFPs, which we discuss.

**Introduction. The GFP Chromophore Motif.**

The impact of the discovery and development of the Green Fluorescent Protein (GFP) on biosciences can hardly be overstated.[1] The GFP chromophore is a *p*-hydroxybenzylidene-imidazolinone (HBI) motif (Scheme 1), which supports multiple protonation states.[2-5] *Aequorea victoria* GFP (*av*GFP), and several variants, display dual absorption maxima near 395 nm (**A** band) and 475 nm (**B** band). The **A** band is normally assigned to the neutral (phenolic) form of HBI while the **B** band is assigned to the anionic (phenolate) form. This assignment is integral to the mechanism of a well-known excited-state proton transfer (ESPT) reaction following excitation into **A** in *av*GFP and variants.[5] This reaction explains why excitation into both **A** and **B** yields the same characteristic green steady-state fluorescence.[6-8] This reaction can be monitored using time-resolved fluorescence and time-resolved vibrational spectroscopies.[6-12]

**Scheme 1.** HBI and Parent Symmetric Dyes

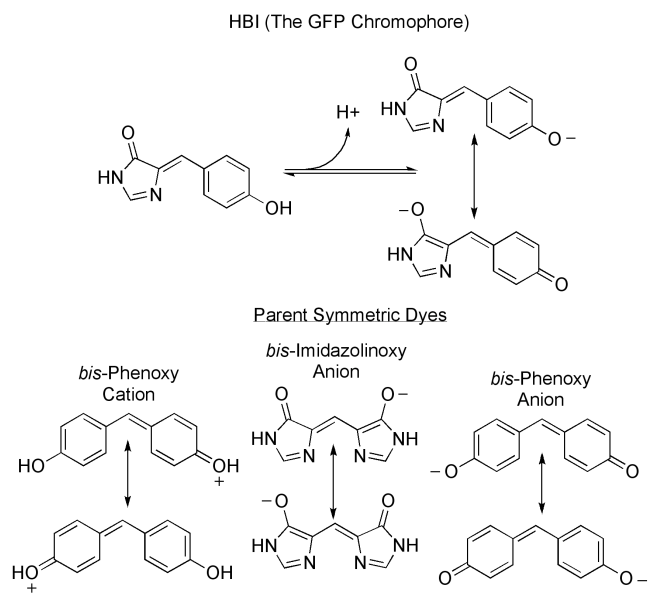

HBI is a *monomethine dye* – specifically, a monomethine oxonol dye.[13] Methine is a limit of sp$^2$ carbon where the Lewis octet rule is satisfied by participation in a charge-transfer resonance (as in Scheme 1). Scheme 1 suggests that HBI anion has this property. Neutral HBI does not; protonation biases one structure over the other, detuning the resonance and yielding olefinic structure with definite bond alternation and charge localization.

In this paper, we will examine a consequence that emerges when HBI is considered as a methine dye ("Brooker dye", see below). Specifically, we will examine the possibility of a higher stationary electronic state with a specific valence structure, which is predicted by the theory of such dyes. The paper will proceed as follows. First, we will review the structure of a particular form of the resonance theory of methine dyes, in the form advocated by Brooker[14] and by Platt[15], paying specific attention to a key component of that theory - the definition of a basicity scale for dye terminal groups. We will argue that the state space originally intended for the theory is not consistent with the definition of such a scale, and suggest a revised state space that fixes this problem. We will show that there are self-consistent solutions to the two- and three-state-averaged CASSCF problems[16] for the dyes in Scheme 1 that have the structure of the revised state space. We invoke these solutions to argue that a higher stationary state, predicted by the heuristic resonance theory, should exist for the anionic chromophore. We calculate its excitation energy, finding it lies just below the first excitation energy of the neutral chromophore. This has important implications for photochemical mechanism in GFPs. We conclude with suggestions for experimental verification.

**The Resonance Color Theory and the Brooker Basicity Scale**

The middle of the twentieth century saw considerable developments in the problem of color and constitution in methine dyes.[14,15,17,18] A major impetus for the theoretical development was the experimental work of Brooker and coworkers, who synthesized and characterized the spectra of a great many methine dyes (hence "Brooker dye" in the title).[14]

Brooker's work highlighted empirical rules for the optical properties of methine dyes. One such rule, the "Deviation Rule", states that the maximum absorbance wavelength of an asymmetric dye (containing different terminal heterocyclic groups, e.g. the GFP chromophore in Scheme 1), is no redder than the mean absorbance wavelengths of the symmetric "parent" dyes (generated by pairing each of the groups with a copy of itself).[14,15] This red limit will be achieved for asymmetric dyes where the "basicities" of the terminal groups are equal. If they are not equal, the absorbance will deviate from the mean by a blue shift that increases as the basicity difference between the terminal groups. In this context, the *basicity* of a group refers to the negative of the energy required to remove *one electron* from the reduced state of the terminal group (as in Scheme 1).

Brooker's work was interpreted in terms of a heuristic theory based on the resonance of Lewis structures (e.g. Scheme 1). Within this theory, the optical excitation arises from a dressed interaction between two resonating *extreme* structures (such as shown in Scheme 1 for the anion). The extreme structures, which are isoenergetic for a symmetric dye, are effectively coupled through *intermediate* structures with charge localized on the bridge (Figure 1).[17] For a monomethine dye, there is only one intermediate structure. For a symmetric dye, only the even parity combination of extreme structures interacts with the intermediate structure. The resulting gap between even and odd parity combinations of extreme states will be optically intense by simple dipole length arguments.[19,20] A second excitation can also emerge, and should be dark by similar arguments.[17,20]

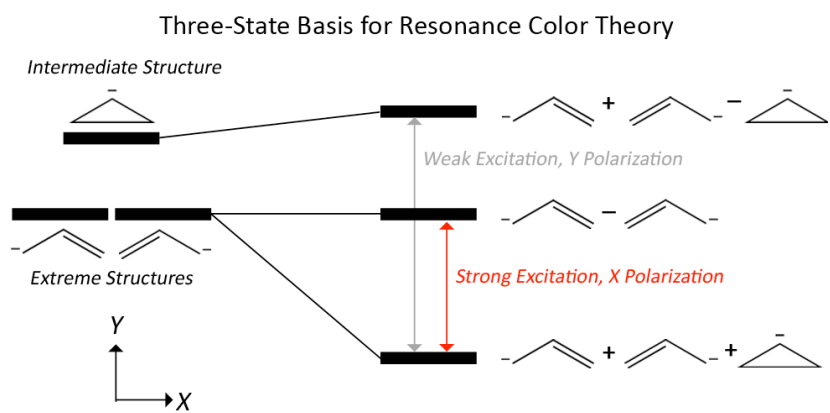

*Figure 1.* Three-state resonance color theory for a symmetric monomethine dye, using allyl anion as an example. Isoenergetic "extreme" Lewis structures, distinguished by bond alternation and charge location, are split by their interaction with a third, higher, structure. The interaction among the three yields a strong optical excitation and a dark higher-energy excitation with orthogonal polarization. The lower state is brighter than the higher state because the vector characterising the net charge motion in the resonance is larger, as the distance between the rings compared to the height of the bridge.

In an asymmetric dye, the extreme structures will have different energy.[15,17] This leads to a residual splitting in the absence of interaction, and contributes an additional blue shift when the interaction is taken into account.[14,15,17] This is the physical origin of the deviation rule. The extreme structures have the same number of bridge bonds, so the splitting is dominated by the basicity difference between the terminal groups. Platt demonstrated that Brooker's deviation data could, within the experimental error, be reconstructed using a single basicity for each of the terminal groups, independent of the dyes with which they were conjugated.[15] The Brooker basicity scale is correlated with other chemical basicity scales, for example the Hammet $\sigma_R$ scale.(c.f. p.250 of [21])

**Prediction of a Higher Excited State in the Resonance Color Theory**

The intermediate structure is a *postulate* of the resonance color theory, which, under certain constraints, leads to the *prediction* of a higher valence stationary state. The postulate is required because the bare interaction of the rings is insufficient to explain the first optical transition.[22] Additional effective interaction through the intermediate structures accounts for the additional splitting. The participation of the intermediate states in the first excitation is marginal, and for the purposes of describing this excitation it can be replaced by a suitable effective potential.[17]

The theory does not require that the intermediate structure give rise to a third stationary state. It may not be stationary, due to strong mixing with the ionic states that we have left implicit in Scheme 1, but which are required in any orthogonal valence bond theory.(c.f. Section 3.3 of [23]) The higher covalent state may not be a good approximation to any one of the stationary states of the full space. If there is such a stationary state, however, then the postulate also constitutes a prediction. We *naively* suggest that is most likely near resonance, because we intuitively expect that fluctuations from the intermediate state into the left and right polarized ionic states should be about equal. We provide support for this intuition below, based on the ability to obtain two- and three-state self consistent field solutions.

The higher excited state is expected to be dark because it is a charge resonance excitation between structures where charge is delocalized over both rings and a structure where it is localized on the bridge.

The transition dipole can be approximated with the vector characterising the net charge motion.[19,20] The distance between the rings (*x* direction, see Fig 1) is larger than the height of the bridge (*y* direction, Fig. 1), so the first excitation is brighter than the second. If an appropriate substituent is appended to the methine, such that the electron can delocalize onto it and lengthen the dipole, then the second transition may be detectable in the linear spectra. This is the origin of the *y band* observed in the dyes Malachite Green and Benzaurin.(c.f. pp.250-260 of [21]) Linear excitation into this state is well-known and fluorescence from it has been characterized.[24,25] Its characteristics are consistent with the higher excitation predicted by the resonance color theory.

For an unsubstituted methine, we would not expect the higher state to be optically intense. If it appears in linear spectra, than it should be recognizable by its polarization.(c.f. p.252 of [21]) It may be detectable by higher-order spectroscopies, or by other signatures. If it correlates with different photochemical products, then wavelength-dependent yields may arise.

**A Reappraisal of the State Space in the Resonance Color Theory**

We argue that representing the color theory in a state space of Lewis structures (such as Scheme 1) is *formally inconsistent* with the content of the theory, as advocated by Brooker and by Platt.[14,15] The problem is that the Lewis structures (interpreted as electron pair products over an atomic basis) contain *too much of the wrong kind of information*. "Basicity" refers to an energy associated with a *one-electron* process, while Lewis structures such as in Scheme 1 are rich in *pair information*.

There are two ways to look at the problem. The first notes that the Lewis structure representation implies a basis of atomic orbitals (which may be orthogonal). If the energies required to remove an electron from each of these is different, than there will be dispersion in the basicity *even for a single dye prepared in a given Lewis structure*. This is because there is more than one way to remove an electron from the group, and there will be amplitudes for all possible ways. The specification of *group basicities* (as is the Brooker-Platt theory) is not compatible with such a detailed description of the rings.

Another perspective notes that, in general, one-electron density matrix elements and two-electron density matrix elements do not commute, nor do they form a closed Lie algebra under commutation.[26]

Therefore, *quantum uncertainty prevents simultaneous, precise specification of the basicity of a group and bonding within the group*. Again, we conclude that there must be dispersion in the group basicities for a single dye prepared in a given structure. Alternatively, the *assertion* of precise basicities implies the presence of *many* contributing structures (many more than suggested by Scheme 1).

The complications above are already apparent from Brooker's work.[14] To interpret the basicities he obtained for the terminal groups in his set, Brooker had to resort to invoking additional structures beyond the canonical pair (c.f. Fig. 16 of [14]). This is not surprising because, again, one and two electron observables do not commute, nor together span a closed Lie algebra. Representing one-electron properties in a basis of products of pair states generally requires long expansions, up to the dimension of the enclosing *complete* space. There need be no natural truncation scheme that will work for all of a given set of cases.

We can circumvent these problems, without fundamentally altering the theory, if we invoke a different representation. Specifically, we advocate the position that a three-orbital complete active space valence bond representation (such as shown in Fig. 2, see Section 9.2.3 of [23]) is more consistent with the content of the resonance color theory. This representation is built from a one-electron active space of three localized orbitals on the two rings and the methine itself. Arranging four electrons over these orbitals in all possible ways gives rise to three "covalent" configurations and three "ionic" configurations, as shown. This representation is *consistent* with the precise definition of group basicities, because we can associate the basicity unambiguously with the difference of energies that differ by one electron in the associated orbital. This definition will be unambiguous so long as the mapping between the orbitals and the rings is unambiguous. Also, we point out that the many-electron space generated by the four-electron, three-orbital ansatz is capable of spanning three *perfect pairing states* (covalent-ionic pairs, see pp.240-242 of [23]). This is enough to index the structures in the canonical resonance of Scheme 1, which, due to the *collective* nature of the charge and bond order alternation, contains information equivalent to *two* perfect pairs.

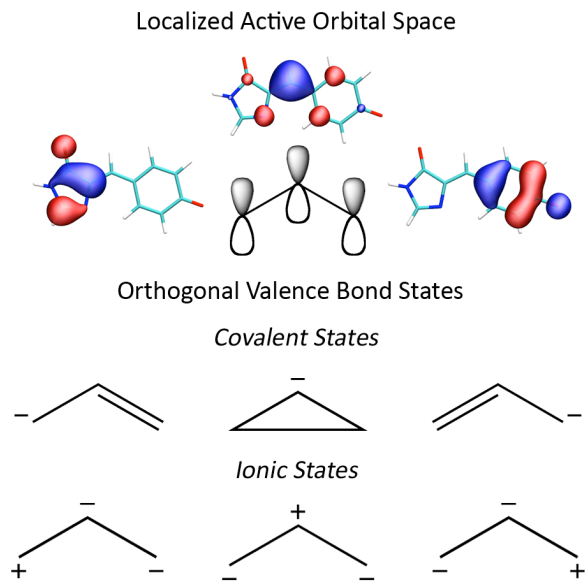

*Figure 2.* A complete active space valence bond (CASVB) representation that is consistent with Brooker's basicity scale. The active orbital space is spanned by localized orbitals on the heterocyclic termini and the bridge. Distributing four electrons among the three orbitals yields six singlet configurations with a natural valence-bond interpretation. There are three covalent and three ionic configurations, sufficient to span three orthogonal perfect pairs.

Dyes such as in Fig. 1 are isoelectronic with *odd alternant systems*.[18,21] When the π orbitals of odd alternant systems are paired off into bonding orbital (BO) and antibonding orbital (ABO) combinations, there will be an odd nonbonding orbital (NBO) left over. In dyes with an even number of π pairs (such as Scheme 1 and Fig. 1), this orbital will be doubly occupied.[18] The first optical transition is modeled as a single excitation from the NBO to the lowest ABO. In dyes with a linear "*y* band" transition (such as Malachite Green and Benzaurin), the higher excitation is modeled as a transition from the highest occupied BO to the lowest occupied ABO.(c.f. p.252 of [21]) Therefore, these theories lead to a complete active space with the same rank that we are proposing, but by a distinctly different chain of arguments.

**Self-Consistent Model Spaces for the Resonance Color Theory**

We have found that there is a solution of the two-state averaged, four electron-three orbital complete active space self-consistent field[16] (SA2-CAS(4,3)) problem, with the form of Fig. 2, which can be readily obtained for several dyes from the "diarylmethane" class (e.g. Michler's Hydrol Blue, Malachite

Green, Auramine O, Benzaurin, etc.). If the active space orbitals are localized via the Foster-Boys[27] criterion, the orbitals on the rings have a transferrable structure, and maintain their form when the group in question is paired with different terminal groups in a set of dye structures. We show this in Figure 3. *This is exactly the transferability pattern presumed in the Brooker basicity scale*.[14,15]

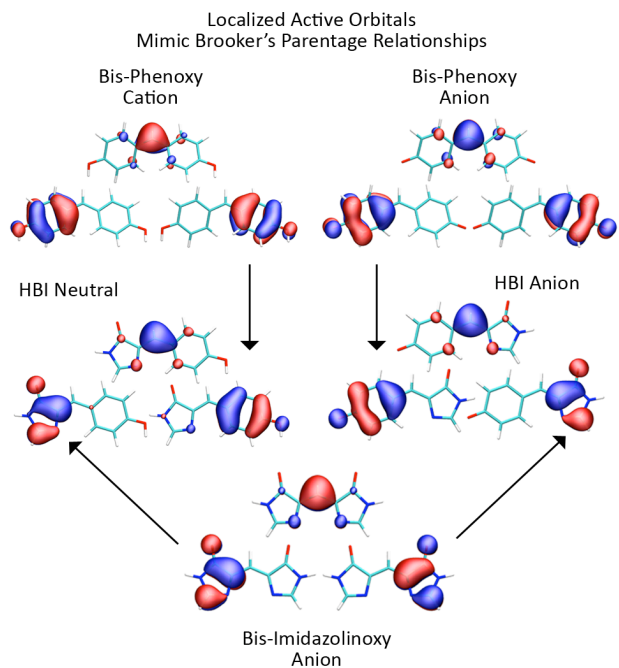

*Figure 3.* Localized active space orbitals from a family of two-state-averaged complete active space self-consistent field solutions for HBI and its symmetric parents. The localized active orbitals have an transferrable structure, which mimics the parentage relationships implied by Brooker's basicity scale.

We have recently applied this ansatz to the excitations of a variety of GFP chromophore protonation states, as well as the corresponding symmetric parent dyes, and found that the calculated excitation energies can be accurately reproduced by a set of basicity parameters specific to each terminal group, and independent of the groups to which they are paired.[28] *This is the same relationship that Platt established for Brooker's data*.[15] An important result that emerged is that HBI anion is *resonant*, with an extremely small Brooker deviation. Neutral HBI has a much larger deviation – and so is non-resonant. *This is our first principal result*. We summarize it in Figure 4. The relevant excitations were calculated using multi-state, multireference perturbation theory[29] (MS-MRPT2) on the SA2-CAS(4,3) reference space with a cc-pvdz basis set[30] at MP2-optimized[31] geometries, with Molpro.[32] A similar solution for the SA2-CAS(2,3) problem can be obtained for dyes with an even number of π electron pairs (for example, monomethine cyanines).[33]

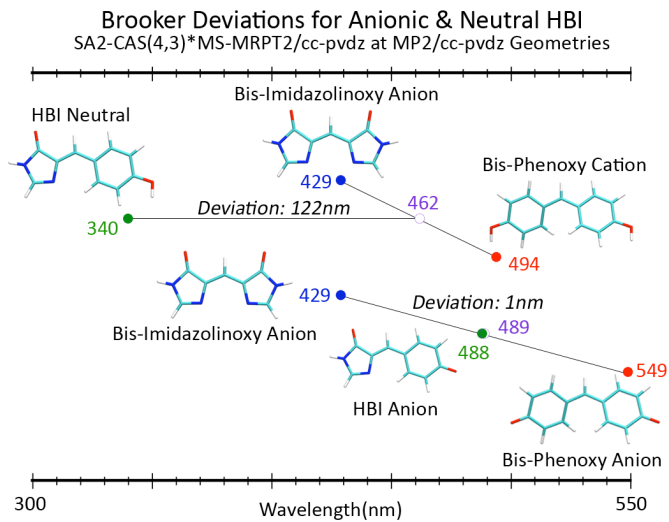

*Figure 4.* Calculated Brooker deviations for neutral (top) and anionic (bottom) HBI. Anionic HBI has a small Brooker deviation, demonstrating that it is near resonance. Neutral HBI has a large deviation, demonstrating that it is not.

There is *also* a *three-state* averaged solution (SA3-CAS(4,3)), with the form of Fig. 2, for the anionic form of HBI and its parent symmetric dyes. The Boys-localized active orbitals in this case are visually indistinguishable from the two-state solution. We have only been able to obtain the three-state self-consistent solution for dyes with a very small Brooker deviation, such as HBI anion. For nonresonant dyes, such as the phenolic neutral form, a SA3-CAS(4,3) procedure does not converge to a solution space with the structure of Fig. 2. The configuration interaction vector for the third state of the anion is dominated by the bridge-charged covalent configuration (Fig. 2). Its transition dipole is two orders of magnitude smaller than the first excitation, and is nearly orthogonal to the first transition dipole (see Fig. 5). These facts support its identification with the higher state predicted by the resonance color theory.

The existence of the three-state averaged self-consistent solution for the anionic form of HBI, but not for non-resonant forms, supports our initially intuitive proposal that the resonance theory predicts a second stationary excited state for resonant dyes. *This is the second principal result of this paper*.

Using MS-MRPT2 on the SA3-CAS(4,3) reference space for HBI anion, we can calculate the excitation energy of the higher state. In Figure 5, we display these, alongside two-state excitation energies obtained for the neutral and anionic forms. The first excitation energy of the anion does not change much between the two- and three-state models, consistent with the assertion that the intermediate structure in the color theory does not play a direct role in the first excitation (and showing that inclusion of the second excited state does not degrade the description of the first). The second excitation of the anion occurs at energies relevant to GFP photochemistry, and lies just to the red of the first excited state of neutral HBI. This suggests that there may be a dark anionic excited state underlying the red edge of the **A** band in GFPs, or between the **A** and **B** bands. *This is our third principal result*.

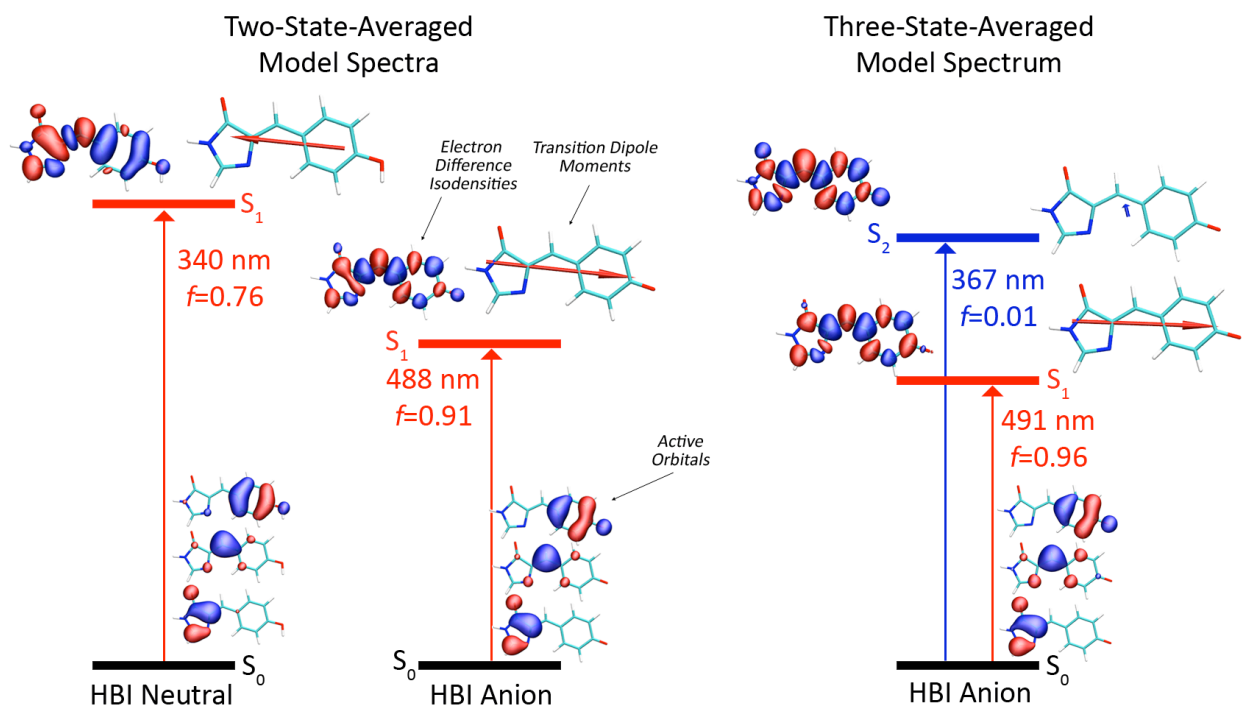

*Figure 5.* Excitation energies, oscillator strengths, transition dipole moments, and electron difference isodensity surfaces for two-state averaged model excitations of neutral and anionic HBI (left) and three-state-averaged excitations of anionic HBI (right). The excitation energy of the second excited state of the anion (3-state model) is closely below the first excited state of the neutral form (2-state model). Note that switching from a two- to a three-state model for the anion has no significant effect on the active space orbitals, nor the excitation energy and oscillator strength of the first excitation.

We have chosen our ansatz based upon formal similarity to the resonance color theory (which was, in turn, distilled from a large body of empirical data), rather than by testing its convergence numerically against alternative approaches to the Born-Oppenheimer electronic structure problem. The latter approach has been taken, for example, by Epifanofsky and coworkers, who have provided state-averaged natural orbitals and occupation numbers from a SA3-CAS(14,12) calculation on a dimethylated form of HBI (HBDI).[34] Their occupation numbers support our choice of active space, because large changes in the occupation border an analogous three-orbital space (see the Supplement to [34]). State-averaged natural orbitals and occupations for our data are included in an online Supplement. The first excitation we obtain for HBI anion is consistent with the absorption of HBDI in an ion trap[35], and with the reddest absorbance recorded for the same compound in a range of solvents.[35,36] The excitation we obtain for the bis-phenoxy parent dyes is similar to the solution state absorbance of Benzaurin anions and cations.(c.f. p.258 of [21]) The absorbance of neutral HBI that we calculate is somewhat bluer than the absorbance of neutral HBDI in a range of solvents[36], with the absorption of neutral models in an ion ring and with higher-rank active space calculations.[37] This does not necessarily discount our result, because the excitation we obtain for the higher anionic state is still somewhat redder than the neutral excitation.

**Implications of a Higher Excitation in Anionic GFP Chromophores**

There are recent experimental results for GFPs whose interpretation may be affected by the presence of a second anionic excitation, lying either under the **A** band or between the **B** and **A** bands. We will highlight some of these.

Excitation into the **A** band of the S65T/H148D mutant of *av*GFP at pH 5.6 yields green emission on an ultrafast (< 100 fs) timescales.[8,12] If this is to be interpreted as ESPT, then there should be an anionic state to accept population on these timescales. A higher anionic state would suffice, because $S_2 \rightarrow S_1$ internal conversion in similar dyes is ultrafast.[24,25] Transient vibrational spectra are consistent with direct excitation of an anionic state or with neutral and anionic electronic states coupled to a common vibration.[11,12]  Stark effect spectra of the mutant at pH 5.6 show anomalous behavior consistent with a hidden state.[8]

Dronpa is a reversibly photoswitching fluorescent protein (RSFP) from the GFP family.[38]  It is fluorescent (ON) in its native state.  The absorbance spectrum in ON is dominated by a **B** band.  Irradiation into **B** yields a non-fluorescent state (OFF) with increased **A** absorbance.  ON returns thermally, but can be immediately regenerated by excitation into **A**.  **A** band populations created by photoswitching and pH manipulation are distinct.[38]  They show different optical nonlinearities, suggesting the distinction may be electronic.[39]  Different optical nonlinearities suggest disjoint populations with different detunings from resonance.[40]  Both photoisomerization and ESPT are implicated in the switching mechanism.[38,41]  Our previous work suggests that the third state may lead to a different photoisomerization dynamics, and this may have implications for the switching mechanism.[33,42]  Correlations between photoisomerization and ESPT phenomena emerge naturally from general considerations of the methine electronic structure.[43]  Qualitatively similar photoswitching behavior has been observed via single molecule spectroscopic studies of GFP variants.[44]

The presence of an anionic state at **A** band wavelengths would be important to consider in formulation of mechanistic hypotheses.  Further work is needed to flesh out these possibilities.

**Strategies for Experimental Investigation**

The resonance color theory (as well as our calculations) predicts that the second excitation should be darker than the first. Oscillator strengths for the second excitation in HBI anion are 2 or more orders of magnitude lower than the first (Fig. 5).  The theory predicts a different polarization for the higher state, as we observe in the calculations (Fig. 5).  This is similar to the case of Malachite Green and Benzaurin, where an $S_0 \rightarrow S_2$ transition dipole is polarized along the axis of the phenyl substituent on the bridge, and orthogonal to the $S_0$-$S_1$ dipole.(c.f. p.252 of [21])  We have not calculated the two-photon absorption (TPA) cross section, but the out-in charge motion characteristic of the excitation suggests enhanced TPA.[45] TPA spectra of HBDI do suggest a hidden excited state.[46] The role of the third state in the photoisomerization reaction suggests excitation to this state may alter the photoisomer yield.[33,42] Excitation-dependence in the distribution of photoisomer yields therefore provides the basis for an indirect test.

The $S_0$-$S_1$ excitation in HBI anion is near the red limit implied by Brooker's deviation rule. Anionic HBI is *resonant*. This result has broader implications for the photobehavior than the suggestion of a third valence state at low energies. It explains, for example, the large solvatochromic shifts recorded for HBDI anion.[36] Significant solvatochromism is consistent with high polarizability, as suggested by the resonance condition.[40]

Resonant electronic structure can leave traces in the vibrational spectra, as highlighted by the exaltation of a $b_{2u}$ mode in the $B_{2u}$ excited state of benzene.(c.f. Section 6.10.1 of [23]) We are unaware of similar phenomena having been reported in HBI derivatives or GFPs. Calculated resonance Raman spectra predict a strong signal at 1350cm$^{-1}$ in anionic HBDI, assigned to a delocalized bridge vibration.[47] This feature is suppressed in solution-state spectra of anionic HBDI, which are dominated by a line in the 1500-1600cm$^{-1}$ range.[47] A similar feature emerges when the spectrum is computed in a polarizable continuum, and is correlated with increased bond alternation.[48] Other methine dyes show resonance[49] and non-resonance[50] Raman intensity near 1350cm$^{-1}$, as does the pre-resonance Raman spectrum of a photochromic GFP variant.[47]

**Conclusion**

We have presented quantum chemical evidence that there is a higher valence excitation of the anionic GFP chromophore, which lies closely below the first excitation of the neutral form. The valence character of the state is consistent with a state predicted by the resonance color theory of methine dyes. If our prediction is confirmed, it will have significant implications for the mechanistic interpretation of GFP photophysics. We have highlighted specific experiments that a higher anionic state might explain, and have suggested experimental strategies for further investigation.

**Acknowledgement** This work was supported by the Australian Research Council under Discovery Project DP0877875. Computations were done at the National Computational Infrastructure (NCI) Facility, Canberra, with time provided under Merit Allocation Scheme (MAS) project m03. We thank J.R. Reimers, N.S. Hush and A.N. Tarnovsky for bringing Brooker's work to our attention, as well as S. Boxer, W. Domcke, S. Marder, S. Meech, T. Martínez, M. Prescott, M. Robb, M. Olivucci, G. Groenhof, T. Pullerits, T. Smith, M. Smith, S.C. Smith and R. Jansen-Van Vuuren for helpful discussions.

**Supporting information available** Molecular coordinates (Å), SA-CASSCF and MS-MR-RSPT2 absolute energies (a.u.), SA-CASSCF natural orbital images and occupation numbers, MS-MR-RSPT2 mixing matrices are available upon request.

## References

[1] M. Zimmer. Chem. Soc. Rev. 38 (2009) 2823.

[2] S. Meech. Chem. Soc. Rev. 38 (2009) 2922.

[3] S. Remington. Curr. Op. Struct. Bio. 16 (2006) 714.

[4] J.J. van Thor. Chem. Soc. Rev. 38 (2009) 2935.

[5] P.J. Tonge, S.R. Meech. J. Photochem. Photobiol. A205 (2009) 1.

[6] M. Chattoraj, B. King, G. Bublitz, S. Boxer. Proc. Natl. Acad. Sci. U.S.A. 93 (1996) 8362.

[7] T. McAnaney, E. Park, G. Hanson, S. Remington, S. Boxer. Biochemistry 41 (2002) 15489.

[8] X. Shi, P. Abbyad, X. Shu, K. Kallio, P. Kanchanawong, W. Childs, S. Remington, S. Boxer. Biochemistry 46 (2007) 12014.

[9] D. Stoner-Ma, E.H. Melief, J. Nappa, K.L. Ronayne, P.J. Tonge, S.R. Meech. J. Phys. Chem. B110 (2006) 22009.

[10] D. Stoner-Ma, A.A. Jaye, P. Matousek, M. Towrie, S.R. Meech, P.J. Tonge. J. Am. Chem. Soc. 127 (2005) 2864.

[11] D. Stoner-Ma, A.A. Jaye, K.L. Ronayne, J. Nappa, S.R. Meech, P.J. Tonge. J. Am. Chem. Soc. 130 (2008) 1227.

[12] M. Kondo, I. Heisler, D. Stoner-Ma, P. Tonge, S. Meech. J. Am. Chem. Soc. ASAP (Epub date Nov 16, 2009) (2009).

[13]  H. Berneth, Methine Dyes and Pigments, Ullmann's Encyclopedia of Industrial Chemistry. John Wiley & Sons Inc., 2009.

[14]  L.G.S. Brooker. Rev. Mod. Phys. 14 (1942) 275.

[15]  J. Platt. J. Chem. Phys. 25 (1956) 80.

[16]  H.-J. Werner, W. Meyer. J. Chem. Phys. 74 (1981) 5794.

[17]  W. Moffitt. Proc. Phys. Soc. A63 (1950) 700.

[18]  M. Dewar. J. Chem. Soc. (1950) 2329.

[19]  R. Feynman, R.B. Leighton, M.L. Sands. Quantum Mechanics, Addison-Wesley Publishing Company, Reading, , 1989. p. 10-12.

[20]  R. Mulliken. J. Chem. Phys. 7 (1939) 20.

[21]  J. Griffiths. Colour and Constitution of Organic Molecules, Academic Press Inc., London, 1976

[22]  J. Reimers, N. Hush. Nanotechnology 7 (1996) 417.

[23]  S.S. Shaik, P.C. Hiberty. A Chemist's Guide to Valence Bond Theory, John Wiley & Sons, Inc., Hoboken, NJ, 2008

[24]  M. Yoshizawa, K. Suzuki, A. Kubo, S. Saikan. Chem. Phys. Lett. 290 (1998) 43.

[25]  A. Bhasikuttan, A. Sapre, T. Okada. J. Phys. Chem. A107 (2003) 3030.

[26]  H.J. Lipkin. Lie Groups for Pedestrians, Dover Publications Inc., Mineola, N.Y., 2002. p. 16-17.

[27]  J. Foster, S. Boys. Rev. Mod. Phys. 32 (1960) 300.

[28]  S. Olsen.  (To be published.).

[29]  P. Celani, H.-J. Werner. J. Chem. Phys. 112 (2000) 5546.

[30]  T.H. Dunning. J. Chem. Phys. 90 (1989) 1007.

[31]  A. El Azhary, G. Rauhut, P. Pulay, H.-J. Werner. J. Chem. Phys. 108 (1998) 5185.

[32]  MOLPRO, version 2009.1, a package of ab initio programs, H.-J. Werner, P. J. Knowles, R. Lindh, F. R. Manby, M. Schütz, and others , see http://www.molpro.net.

[33]  S. Olsen, R.H. McKenzie. J. Chem. Phys. 131 (2009) 234306.

[34]  E. Epifanovsky, I. Polyakov, B. Grigorenko, A. Nemukhin, A. Krylov. J. Chem. Theo. Comput. 5 (2009) 1895.

[35]  M. Forbes, R. Jockusch. J. Am. Chem. Soc. 131 (2009) 17038.

[36]  J. Dong, K. Solntsev, L. Tolbert. J. Am. Chem. Soc. 128 (2006) 12038.

[37]  J. Rajput, D.B. Rahbek, L.H. Andersen, T. Rocha-Rinza, O. Christiansen, K.B. Bravaya, A.V. Nemukhin, A.V. Bochenkova, K.M. Solntsev, J. Dong, J. Kowalik, L.M. Tolbert, M.Å. Petersen, M.B. Nielsen. Phys. Chem. Chem. Phys. 11 (2009) 9996.

[38]  S. Habuchi, R. Ando, P. Dedecker, W. Verheijen, H. Mizuno, A. Miyawaki, J. Hofkens. Proc. Natl. Acad. Sci. U.S.A. 102 (2005) 9511.

[39] I. Asselberghs, C. Flors, L. Ferrighi, E. Botek, B.t. Champagne, H. Mizuno, R. Ando, A. Miyawaki, J. Hofkens, M.V.D. Auweraer, K. Clays. J. Am. Chem. Soc. 130 (2008) 15713.

[40] S. Marder, C. Gorman, F. Meyers, J. Perry, G. Bourhill, J. Brédas, B. Pierce. Science 265 (1994) 632.

[41] M. Andresen, A. Stiel, S. Trowitzsch, G. Weber, C. Eggeling, M. Wahl, S. Hell, S. Jakobs. Proc. Natl. Acad. Sci. U.S.A. 104 (2007) 13005.

[42] S. Olsen, R.H. McKenzie. J. Chem. Phys. 130 (2009) 184302.

[43] S. Olsen, K. Lamothe, T.J. Martínez. J. Am. Chem. Soc. (In Press).

[44] W.E. Moerner. J. Chem. Phys. 117 (2002) 10925.

[45] M. Albota, D. Beljonne, J.-L. Brédas, J.E. Ehrlich, J.-Y. Fu, A.A. Heikal, S.E. Hess, T. Kogej, M.D. Levin, S.R. Marder, D. McCord-Maughon, J.W. Perry, H. Röckel, M. Rumi, G. Subramaniam, W.W. Webb, X.-L. Wu, C. Xu. Science 281 (1998) 1653.

[46] H. Hosoi, S. Yamaguchi, H. Mizuno, A. Miyawaki, T. Tahara. J. Phys. Chem. B112 (2008) 2761.

[47] S. Luin, V. Voliani, G. Lanza, R. Bizzarri, R. Nifosì, P. Amat, V. Tozzini, M. Serresi, F. Beltram. J. Am. Chem. Soc. 131 (2008) 96.

[48] P. Altoe, F. Bernardi, M. Garavelli, G. Orlandi. J. Am. Chem. Soc. 127 (2005) 3952.

[49] J.P. Yang, R.H. Callender. J. Raman Spectrosc. 16 (1985) 319.

[50] N.O. McHedlov-Petrosyan, Y.N. Surov, V.A. Trofimov, A.Y. Tsivadze. Theor. Exp. Chem. 26 (1991) 644.